\documentclass[
twocolumn,
preprintnumbers,
 aps,
 floatfix,
 prl,
 nofootinbib,
 superscriptaddress,
 showpacs]{revtex4-1}

\usepackage[utf8]{inputenc}

\usepackage{hyperref}
\usepackage{graphicx}

\usepackage[english]{babel}

\usepackage{float}

\usepackage[printonlyused]{acronym}
\usepackage{subfigure}

\usepackage{latexsym}
\usepackage{bm}
\usepackage{amsmath}
\usepackage{amssymb}
\usepackage{amsthm}

\usepackage{placeins}

\usepackage{gensymb}

\usepackage{color}
\usepackage{slashed}
\usepackage{xcolor}
\usepackage{bbold}

\usepackage{bm,slashed,multirow,
	soul,mathtools,xspace,array,comment}  

\usepackage{psfrag,epsfig,graphicx,graphics}

\usepackage{epsf}
\usepackage{amsfonts}

\def\slashchar#1{\setbox0=\hbox{$#1$}
	\dimen0=\wd0
	\setbox1=\hbox{/} \dimen1=\wd1
	\ifdim\dimen0>\dimen1
	\rlap{\hbox to \dimen0{\hfil/\hfil}}
	#1
	\else
	\rlap{\hbox to \dimen1{\hfil$#1$\hfil}}
	/
	\fi}

\newcommand{\APP}{App.~}

\newcommand{\FIG}{Fig.~}

\newcommand{\SEC}{Sec.~}

\newcommand{\EQ}{Eq.~}
\newcommand{\EQs}{Eqs.~}

\def\tr{{\rm tr}}

\def\bei{\begin{itemize}}
	\def\ei{\end{itemize}}

\def\beeq{\begin{eqnarray}} 
	\def\beqa{\begin{eqnarray}}
		\def\bea{\begin{eqnarray}}
			
			\def\eea{\end{eqnarray}}
		\def\eqa{\end{eqnarray}}
	\def\eeeq{\end{eqnarray}}

\def\eqar{\end{array}}
\def\beqar{\begin{array}}

\def\beas{\begin{eqnarray*}}
	\def\beqas{\begin{eqnarray*}}
		
		\def\eqas{\end{eqnarray*}}
	\def\eeas{\end{eqnarray*}}

\def\beq{\begin{equation}} 
	\def\be{\begin{equation}}
		
		\def\ee{\end{equation}}
	\def\eq{\end{equation}}
\def\eeq{\end{equation}}

\def\beqd{\begin{displaymath}}
\def\eeqd{\end{displaymath}}
\def\eqd{\end{displaymath}}

\def\beeq{\begin{eqnarray}} \def\eeeq{\end{eqnarray}}

\newcommand{\fin}{\end{document}}

\def\fin{\end{document}}

\begin{document}

	\title{Evidence of collinear factorization breaking due to collinear-to-soft Glauber exchanges for a $2 \to 3$ exclusive process at leading twist}
	
	\author{Saad Nabeebaccus}
	\email{saad.nabeebaccus@manchester.ac.uk}
	\affiliation{Universit\'e Paris-Saclay, CNRS/IN2P3, IJCLab, 91405 Orsay, France}
	\affiliation{Department of Physics \& Astronomy, University of Manchester, Manchester M13 9PL, United Kingdom}
	
	\author{Jakob Sch\"onleber}
	\email{jschoenle@bnl.gov}
	\affiliation{Institut f\"ur Theoretische Physik, Universit\"at Regensburg, D-93040 Regensburg, Germany}
 \affiliation{RIKEN BNL Research Center, Brookhaven National Laboratory, Upton, NY 11973, USA}
	
	\author{Lech Szymanowski}
	\email{Lech.Szymanowski@ncbj.gov.pl}
	\affiliation{National Centre for Nuclear Research (NCBJ), 02-093 Warsaw, Poland}
	
	\author{Samuel Wallon}
	\email{Samuel.Wallon@ijclab.in2p3.fr}
	\affiliation{Universit\'e Paris-Saclay, CNRS/IN2P3, IJCLab, 91405 Orsay, France}

	\begin{abstract}
 We exhibit an exclusive process, namely the photoproduction of a $  \pi ^{0}\gamma  $ pair with large invariant mass, which violates collinear factorization. We explicitly demonstrate that this is due to the fact that there exists diagrams with gluon exchange in $t$ channel, contributing at the leading power, for which \textit{Glauber gluons} are \textit{trapped}. This is caused by the pinching of the contour integration of \textit{both} the plus and minus light-cone components of the Glauber gluon momentum. We argue that this leads to the observed ``endpoint-like'' divergence of the convolution integral at leading order and leading power when collinear factorization is na\"ively assumed.
	\end{abstract}
	
	\maketitle

	\section{Introduction}
	
	Various $ 2 \to 3  $ exclusive processes have been studied in order to probe \textit{generalized parton distributions} (GPDs)
\cite{ElBeiyad:2010pji,Boussarie:2016qop,Pedrak:2017cpp,Pedrak:2020mfm,Grocholski:2021man,Grocholski:2022rqj,Duplancic:2018bum,Duplancic:2022ffo,Duplancic:2023kwe,Qiu:2023mrm}. A proof of factorization for a class of such processes was recently derived by Qiu and Yu \cite{Qiu:2022bpq,Qiu:2022pla}. Moreover, the photoproduction of a diphoton pair was explicitly calculated at next-to-leading order (NLO)~\cite{Grocholski:2021man,Grocholski:2022rqj} and shown to be infrared (IR) finite. However, we discovered that the amplitude for the photoproduction of a $  \pi ^{0}\gamma  $ pair with large invariant mass, which is sensitive to both quark and gluon GPDs, has singularities when the double convolution of the coefficient function (hard part) with the GPD and distribution amplitude (DA) of the $  \pi ^{0} $-meson is performed, already at the LO, in the case of a gluon GPD. This is quite unexpected since similar computations for the photoproduction of a $  \pi ^{\pm}\gamma  $ pair \cite{Duplancic:2018bum,Duplancic:2022ffo} and of a $  \rho ^{0,\,\pm}\gamma  $ pair \cite{Boussarie:2016qop,Duplancic:2023kwe}, which only have contributions from quark GPDs, are free of such singularities.

In this Letter, we describe, for the first time, a situation where the \textit{breakpoints} $x= \pm \xi$, between DGLAP and ERBL regions \cite{Collins:1998be}, are responsible for the breakdown of collinear factorization at leading twist. We show that the origin of the above-mentioned singularities is a Glauber gluon exchange between a soft spectator parton from the pion and a collinear spectator parton from the nucleon. This is caused by the pinching of the contour integration of both the plus and minus components of the Glauber gluon momentum. We stress, however, that the  contribution involving the quark GPD poses no problem, and therefore, the proof of factorization of \cite{Qiu:2022bpq,Qiu:2022pla} still holds for cases where the gluon GPD contribution is forbidden \cite{Boussarie:2016qop,Duplancic:2018bum,Duplancic:2022ffo,Duplancic:2023kwe}. On the other hand, a direct consequence of our work in this letter is that other related processes involving gluon GPD contributions, such as the production of a photon pair from $  \pi ^{0} $-nucleon scattering, considered in \cite{Qiu:2022bpq}, also suffer from the same Glauber gluon problem.
 
After introducing some kinematics for the photoproduction of a $\pi^{0}\gamma$ pair, we start by analyzing the region of loop momentum space where partons are strictly collinear, focusing on the particular diagram in \FIG\ref{fig:explicit-2-loop}. We demonstrate both the (standard) collinear pinch and the Glauber pinch of this diagram, and further show that both contribute to the same leading power in the expansion of the amplitude in the hard scale. We argue that the observed endpoint-like singularity when collinear factorization is assumed is directly linked to the Glauber pinch which we have identified. The existence of a Glauber pinch, contributing at the leading power in the hard scale, implies the breakdown of collinear factorization for this process.

	\section{Kinematics and frame choice}
	
	We focus here on the specific case of $  \pi ^{0}\gamma  $ photoproduction:
	\begin{align}
	\gamma (q) + N(p_{N})	 \to \gamma (q') +  \pi ^{0}(p_{\pi}) + N'(p_{N'})\;,
	\label{eq:process-def}
\end{align}
with $q^2 = q'^2 = 0$, $ p_N^2 = p_{N'}^2 = m_N^2$, and $p_{\pi}^2 = m_{\pi}^2$, where $m_N$ is the nucleon mass and $m_{\pi}$ is the pion mass, and $P =1/2\, (p_N + p_{N'})$, $\Delta = p_{N'} - p_N$ and $t = \Delta^2$.

The kinematic restriction where factorization of the process in \EQ\eqref{eq:process-def} is expected to hold, according to \cite{Qiu:2022pla}, is that in the center-of-mass (CM) frame with respect to the momenta $\Delta$ and $q$, the transverse components of $q'$ and $p_{\pi}$ are much larger than $\sqrt{|t|}, m_{\pi}, m_{N}$. Denoting the generic hard scale by $Q^2 \sim |q_{\perp}'^2|, |p_{\pi,\perp}^2|$ in the CM frame w.r.t. $\Delta$ and $q$, and the generic small scale by $\lambda Q \sim \sqrt{|t|}, m_{\pi}, m_N, \Lambda_{\mathrm{QCD}}$, the condition becomes $\lambda \ll 1$ simply.

We use light-cone coordinates with respect to two light-like vectors $n^{\mu} = 1/\sqrt{2}\, (1,0,0,-1),$ $ \bar n^{\mu} = 1/\sqrt{2}\, (1,0,0,1)$ without loss of generality. For a generic vector $V$, we define $V^+ = n \cdot V,~ V^- = \bar n \cdot V$ and $V_{\perp}^\mu = V^\mu - V^+ \bar n^\mu - V^- n^\mu$. We will commonly denote a vector $V$ in terms of its light-cone components by $V = (V^+, V^-, V_{\perp})$. Furthermore, we define the following scalings\footnote{By scaling, we mean an extended version of the usual asymptotic equivalence, in the sense that $f(x) \sim g(x)$ as $x \rightarrow 0$ if $f(x)/g(x) \rightarrow \text{const.}$ as $x \rightarrow 0$. The limit $x \rightarrow 0$ is omitted from notation and usually clear from the context.} for a generic momentum $l$,
\begin{align}
    l &\sim Q (1, \lambda^2, \lambda) \qquad &&\bar{n}\textrm{-coll.},\\
    l &\sim Q ( \lambda^2,1, \lambda) \qquad &&{n}\textrm{-coll.},\\
    l &\sim Q (\lambda^2, \lambda^2, \lambda^2) \qquad &&\textrm{ultrasoft},\\
    l &\sim Q (\lambda, \lambda, \lambda) \qquad &&\textrm{soft},\\
    \label{eq:coll-to-coll-Glauber}
    l &\sim Q (\lambda^2, \lambda^2, \lambda) \qquad &&\bar{n}\textrm{-coll. to }n\textrm{-coll. Glauber},\\
    l &\sim Q (\lambda,\lambda^2, \lambda) \qquad &&\bar{n}\textrm{-coll. to soft Glauber}.
    \label{eq:nbar-coll-to-soft-Glauber}
   \end{align}
In particular, while we show explicitly two relevant Glauber scalings, we note that a general Glauber scaling corresponds to any momentum $l$ where $l^+ l^- \ll |l_{\perp}|^2$.

For our purposes, it will be more convenient to consider the CM frame with respect to $\Delta$ and $p_{\pi}$. Our choice of frame is defined by $\Delta_{\perp} = p_{\pi,\perp} = 0$. The skewness is defined as $\xi = - \frac{\Delta^+}{2P^+}$. The condition of small $t$ implies that $p_N, p_{N'}$ can be viewed as approximately collinear in the $\bar n$ direction, so that
\begin{align}
p_N, p_{N'}, \Delta, P \sim (1, \lambda^2, \lambda)Q, \qquad p_{\pi} \sim (\lambda^2, 1, \lambda)Q.
\label{eq:external-scalings}
\end{align}
All the momentum components of the two photons are of order $Q$ in this frame, $q, q' \sim (1,1,1) Q$, with the condition that they are real.\footnote{The photons can also be quasi-real, with $q^2 \sim q'^2 \sim \lambda^2$. This does not spoil the arguments in this letter.} Therefore, one should keep in mind that one can have singularities when virtual particles become collinear to $q$ or $q'$. Since the photons are physical particles, their energies, which are the sum of the plus- and minus-momenta, are positive. Together with the on-shell condition, this implies that $q^+, q^-, q'^+ ,q'^- > 0$.

Throughout this letter, we will set $Q=1$ for notational simplicity.
	
\section{Scalings and pinches}

Due to the presence of a hard scale $Q$, the process in \EQ\eqref{eq:process-def} actually involves a hard scattering at the \textit{partonic} level. These partons are part of loops that connect the hard scattering sub-process to non-perturbative correlators that describe the long-distance physics. The next step is to identify the regions of loop momenta for the partons which give the dominant contribution to the amplitude of the process under study in the limit $Q \to \infty$. This can be expressed as ${\cal A} = \sum_{\alpha}f_{\alpha}\lambda^{\alpha}$, 
and it is the first term in this expansion that we want to obtain. This requires the determination of the IR singularities that correspond to specific configurations of the parton loop momenta, which are ``trapped'' by poles of internal propagators. A given component of loop momentum is said to be \textit{trapped}, or equivalently \textit{pinched}, if there are poles parametrically close to each other above and below the contour integration along the real axis.

While for 1-dimensional contours in the complex plane, it is easy to identify the trapping based on the geometric visualization of the possible deformations, this intuitive picture fails already in a 2-dimensional space embedded in $\mathbb C^2$. For standard loop integrals in $d$ dimensions, a key tool to identify such pinches is provided by the Landau equations \cite{Landau:1959fi}, which give the pinch singular surfaces (PSSs). One standard PSS is the collinear pinch, where the partons have large momentum components only in the direction of the involved external hadrons. Collinear factorisation is valid for a process at a given power, if the only relevant\footnote{There may also be (ultra)soft PSSs which contribute at leading power. They typically lead to vacuum matrix elements of Wilson lines, and for many processes simply give unity.} IR singularities are due to the collinear PSSs. This then implies that the amplitude of the process can be described as a convolution between the hard partonic sub-process and universal distributions (DAs, GPDs) over the longitudinal momentum fractions of the partons linking them.

In fact, the leading power behaviour of the neighborhood of such a PSS arises because the integrand becomes very large so that, although the volume of the phase space region may be small, the region still gives a large contribution to the whole integral.

An important subtle point is that the Landau conditions strictly give exact pinches. For example, if an external particle (e.g. meson) has exactly a four-momentum squared of zero, then the Landau conditions predict an exact pinch for the corresponding partonic loop momentum when it is collinear, and when it is soft. In practice, however, external particles have non-zero four-momentum squared, for instance of order $\lambda^2$. The poles are then separated by a non-zero distance of $O(\lambda)$, which means that the pinch is now \textit{approximate}. However, the loop momentum components are still restricted to have certain maximum sizes, due the approximate PSSs limiting possible contour deformations.

An immediate consequence of the Landau condition is that when massless internal particles are involved, the ``generic'' soft pinch is always exact, independently of the external particles, and appears when the momentum of a massless propagator is zero. However, this might not always lead to an IR singularity because of numerator and momentum space volume factors that may suppress this region. Whether there is indeed an IR singularity requires an analysis on a case-by-case basis. In particular, if massless internal particles are involved, one should always consider both the soft and ultrasoft scalings of internal loop momenta, since they are always pinched in the broad sense. However, one can argue that for exclusive processes, the ultrasoft modes are unphysical due to confinement, since their wavelength $\sim 1/\lambda^2$ is larger than the typical size of the target hadron $\sim {1}/{\lambda}$. For this reason, one usually considers the soft region, as opposed to the ultrasoft region, for exclusive processes. Therefore, we will also focus on the soft scaling in this Letter.

\section{Explicit example}

\begin{figure}
    \centering
    \includegraphics[width=4.2cm]{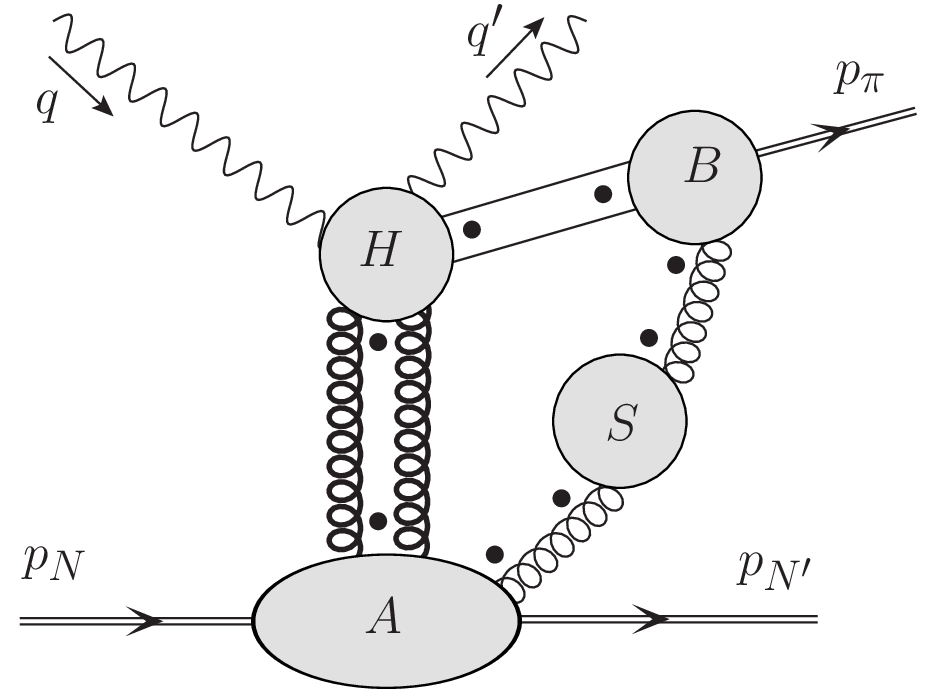}
    \hfill
    \includegraphics[width=4.2cm]{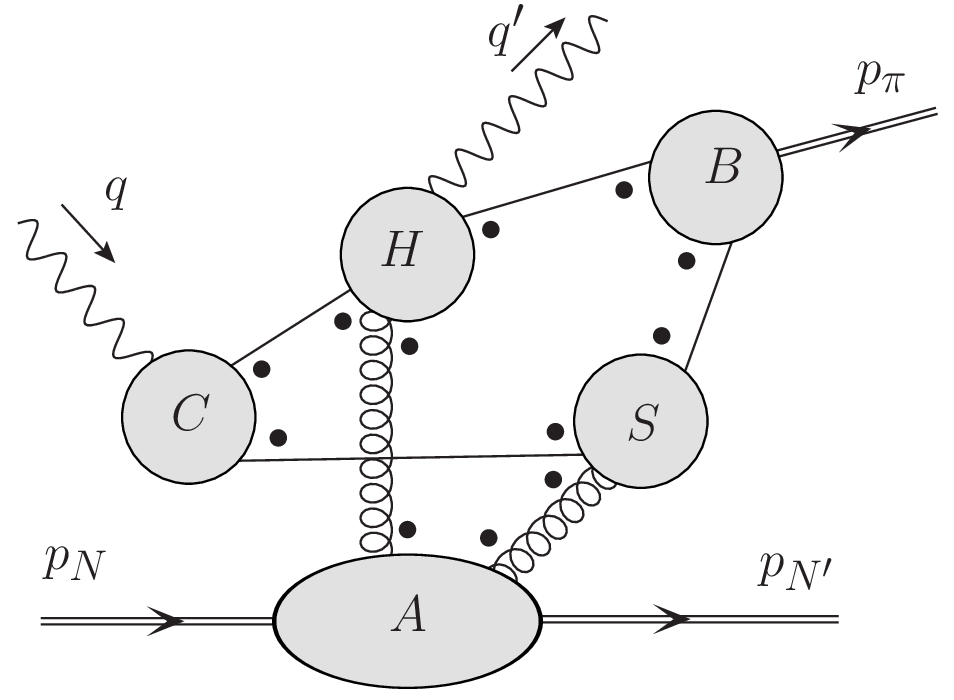}
    (a)\hspace{4.1cm}(b)\hspace{0.4cm}
    \caption{Reduced graphs corresponding to two particular regions of loop momenta for our process. $A,B,C$ denote collinear subgraphs (in different directions), $H$ denotes the hard subgraph and $S$ denotes a (ultra)soft subgraph. Dots beside the lines denote an arbitrary number of scalar-polarized gluons.
    (a): Standard collinear region with two transversely-polarized collinear gluon exchanges (shown in bold) between $A$ and $H$. (b): A region corresponding to a soft gluon exchange between the (ultra)soft spectator parton of the incoming photon ($C$)/outgoing pion ($B$) going through $S$, and a collinear spectator parton from the nucleon sector ($A$).}
    \label{fig:subset-reduced-diagrams}
\end{figure}

\begin{figure}
		\centering
		\includegraphics[width=9cm]{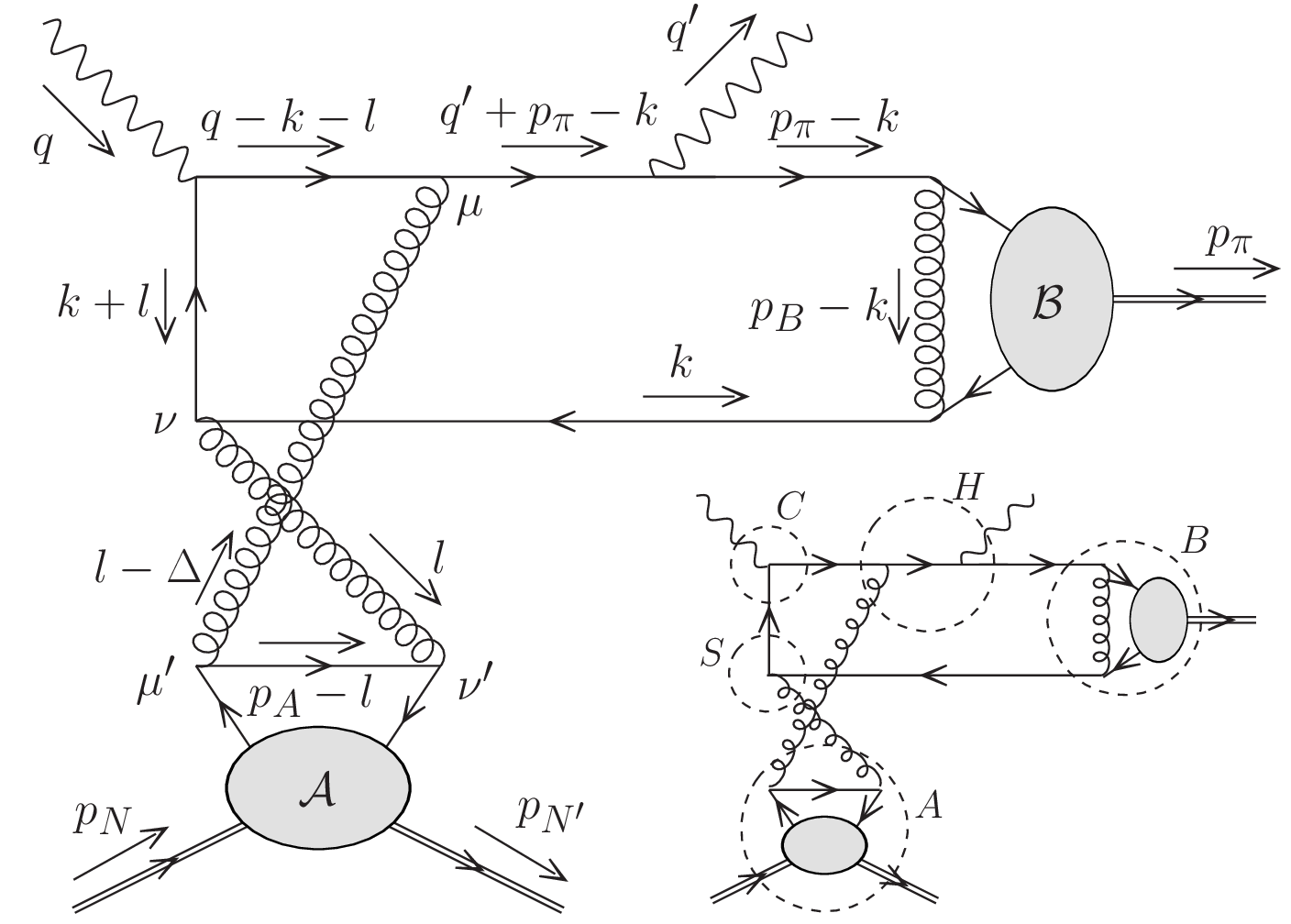}
		\caption{An explicit 2-loop example graph for our process, where the Glauber pinch occurs for the loop momentum $l$. The bottom right graph indicates the mapping to \FIG\ref{fig:subset-reduced-diagrams}(b) for the Glauber region discussed in the main text.}
		\label{fig:explicit-2-loop}
	\end{figure}

It is possible to identify regions of partonic loop momenta which give dominant contributions to the amplitude using Libby-Sterman power counting rules \cite{Libby:1978qf,Sterman:1978bi,Sterman:1978bj,Collins:2011zzd}. We discuss this in detail in \cite{Nabeebaccus:2023rzr}. It is instructive to focus on the two regions shown in \FIG\ref{fig:subset-reduced-diagrams}. The leading power contribution, factorized in terms of the gluon GPD, is shown in \FIG\ref{fig:subset-reduced-diagrams}(a). The corresponding region is such that the two gluons attaching the $A$ to the $H$ subgraphs are strictly collinear to the nucleon. On the other hand, the reduced graph in \FIG\ref{fig:subset-reduced-diagrams}(b) corresponds to an ``endpoint'' contribution, where one of the gluon becomes (ultra)soft. 
We note that the Libby-Sterman power counting rule applies to specific momentum configurations where the internal particles can have collinear, anti-collinear or ultrasoft scalings. Nevertheless, we use \FIG\ref{fig:subset-reduced-diagrams}(b), going beyond the aforementioned specific momentum configurations (in particular, Glauber and soft scalings), and perform the power counting analysis explicitly on a specific two-loop example, shown in \FIG\ref{fig:explicit-2-loop}. 

The loop momenta $p_A$ and $p_B$ circulate only through the subgraphs ${A}$ and ${B}$ and thus have scalings $(1,\lambda^2,\lambda)$ and $(\lambda^2,1,\lambda)$ respectively. The sub-amplitudes ${\cal A}$ and ${\cal B}$ are defined by their vector component in Dirac space,\footnote{We consider the vector case for simplicity. In reality, one should project onto the axial-vector contribution for the leading twist pion DA for ${\cal B}$.} and include with them the external quark propagators, as well as the measures $d^4 p_A$ and $d^4p_B$ respectively. They can thus be taken to be ${\cal A}^{\mu} \gamma_{\mu}$ and ${\cal B}^{\mu} \gamma_{\mu}$ respectively, where, by dimensional analysis and Lorentz covariance, we have ${\cal A} \sim (1, \lambda^2, \lambda)$ and ${\cal B} \sim (\lambda^3, \lambda, \lambda^2)$ (see Chapter 5.5. of  \cite{Collins:2011zzd}). The extra power of $\lambda$ in ${\cal B}$ is due to the fact that it has one external line less than ${\cal A}$.

In the Feynman gauge, the diagram in \FIG\ref{fig:explicit-2-loop} can be written as
\begin{align}
		\label{eq:2-loop-amplitude}
	\int F_{A}^{ \mu ' \nu '} g_{ \mu  \mu '}g_{ \nu  \nu '} \tr  \left[   {F}_{H}^{ \mu  \nu } {F} _{B}  \right] \;,
	\end{align}
where
	\begin{align}
 \label{eq:FA}
		F_{A}^{ \mu'  \nu '} & =  dl^{-}d^{2}l_{\perp}   \frac{\tr \left[  {\slashed{\cal A}}\gamma ^{ \nu '}  \left( \slashed{p}_A- \slashed{l} \right)  \gamma ^{ \mu '}     \right] }{l^2  \left( l- \Delta  \right)^2  \left( p_A-l \right)^2  } \;, \\[5pt]
		F_{B} & = dk^{+}d^{2}k_{\perp} \left( \frac{ \slashed{k} \slashed{\cal B} \left(  \slashed{p}_{ \pi }- \slashed{k}   \right)   }{k^2  \left( p_{ \pi }-k \right)^2  \left( p_B-k \right)^2  } \right) \;,   \label{eq:FB} \\[5pt]
		F_{H}^{ \mu  \nu }& = dk^{-}dl^{+}  \label{eq:FH} \\ & \times \frac{ \slashed{\epsilon }_{q'}^{*} \left(  \slashed{q}'+ \slashed{p}_{ \pi }- \slashed{k}    \right)\gamma ^{ \mu } \left(  \slashed{q}- \slashed{k}- \slashed{l}    \right) \slashed{\epsilon }_{q} \left(  \slashed{k}+ \slashed{l}   \right) \gamma ^{ \nu }   }{ \left( q'+p_{ \pi }-k \right)^2  \left( q-k-l \right)^2  \left( k+l \right)^2   } \;, \nonumber
	\end{align}
and we have omitted the $i \epsilon$ prescriptions.

 The grouping of the measures of the loop momentum components for $l$ and $k$ has been done in the spirit of collinear factorization. That is, with the collinear scaling in \FIG\ref{fig:subset-reduced-diagrams}(a), $F_A$ can be ``identified'' with the gluon GPD, $F_B$ with the pion DA and $F_H$ with the hard coefficient function.

\subsection{Pinch in the collinear region}

The ``standard'' IR singularity is due to the region where $l$ is $\bar{n}$-collinear and $k$ is $n$-collinear, i.e. $l \sim (1,\lambda^2, \lambda)$ and $k \sim (\lambda^2,1,\lambda)$. The corresponding analysis is done in detail in \cite{Nabeebaccus:2023rzr}. The result is that the propagators $k^2$ and $(p_{\pi}-k)^2$ pinch $k^+ \sim {\cal O}(\lambda^2)$, provided $0<k^-<p_{\pi}^-$. The pinch in $l^-$ is generated by any two propagators in \EQ\eqref{eq:FA}, depending on the size and sign of $l^+$. In particular, the $l^2$-denominator is not required for $\Delta^+<l^+<p_{N'}^+$.

\subsection{Power counting in the collinear region}
\label{sec:-pc-for-coll-region}

The collinear region corresponds to $ k \sim ( \lambda ^2, 1,  \lambda ) $ and $ l \sim (1, \lambda ^2, \lambda ) $. For the sake of power counting, we further project onto the \textit{transverse} polarizations of the gluons, i.e. we pick the indices $  \mu, \mu ', \nu, \nu ' $ in \EQs\eqref{eq:FA} to \eqref{eq:FH} to be transverse. This will be justified below. 
Then, we get\footnote{We separate the factors according to the following scheme
\begin{align*}
F \sim \text{momentum space volume} \times  \frac{\text{numerator}}{\text{denominator}}.
\end{align*}}
	\begin{align}
 \label{eq:collinear-scaling}
F_{A}^{ \mu'_{\perp}\nu'_{\perp}} &\sim  \lambda^4 \frac{ \lambda ^2}{ \lambda ^6}= \lambda ^0\,,\qquad F_{B} \sim  \lambda ^4 \frac{ \lambda ^3}{ \lambda ^6}= \lambda ^1\,,\nonumber\\[5pt]
F_{H}^{ \mu_{\perp}\nu_{\perp} } &\sim  \lambda ^{0}\frac{\lambda ^{0}}{\lambda ^{0}} = \lambda^0\;,
	\end{align}
giving an overall scaling of $\lambda$, which we refer to as the leading power contribution, for the diagram in \FIG\ref{fig:explicit-2-loop}. This corresponds to a contribution to the region shown in \FIG\ref{fig:subset-reduced-diagrams}.

It remains to discuss the superleading terms in the polarization sum of the gluons carrying momenta $l$ and $l-\Delta$. To this end, we decompose
\begin{align} \notag
&F_A^{\mu' \nu'} g_{\mu \mu'} g_{\nu \nu'} F_H^{\mu \nu} 
\\
&= F_A^{++} F_H^{--} + F_A^{+\perp} F_H^{-\perp} + F_A^{\perp +} F_H^{\perp -}\nonumber\\
&+ F_A^{\perp\perp} F_H^{\perp\perp} +F_A^{-+} F_H^{+-} +  F_A^{+-} F_H^{-+} +...\,,
\label{eq:FAFH decomp}
\end{align}
where the first (second) line on the RHS corresponds to superleading (leading) terms. When summing over all graphs at the same order in $\alpha_s$ so that the sum is gauge invariant, cancellations due to corresponding Ward identities (WIs) will suppress those superleading terms down to the leading power $\lambda^0$, which is of the order of $F_A^{\perp\perp} F_H^{\perp \perp}$ that was considered in \EQ\eqref{eq:collinear-scaling}. In the Feynman gauge, the detailed analysis is performed in \cite{Nabeebaccus:2023rzr}. More directly, one can choose the light-cone gauge by replacing
\begin{align}
g_{\mu \mu'} \rightarrow g_{\mu \mu'} - \frac{l_{\mu} n_{\mu'}+l_{\mu'} n_{\mu}}{l^+}\,,
\label{eq:lc gauge}
\end{align}
and similarly for $g_{\nu \nu'}$. Since $l_{\mu} = l^+ \bar n_{\mu} + O(\lambda)$, one finds that the three superleading terms in \EQ\eqref{eq:FAFH decomp} are absent
This in turn implies that in the Feynman gauge, all of the three superleading terms in \EQ\eqref{eq:FAFH decomp} will be at least suppressed to the leading power due to WI cancellations.

\subsection{Pinch in the Glauber region}

\begin{figure}
    \centering
    \includegraphics[width=6cm]{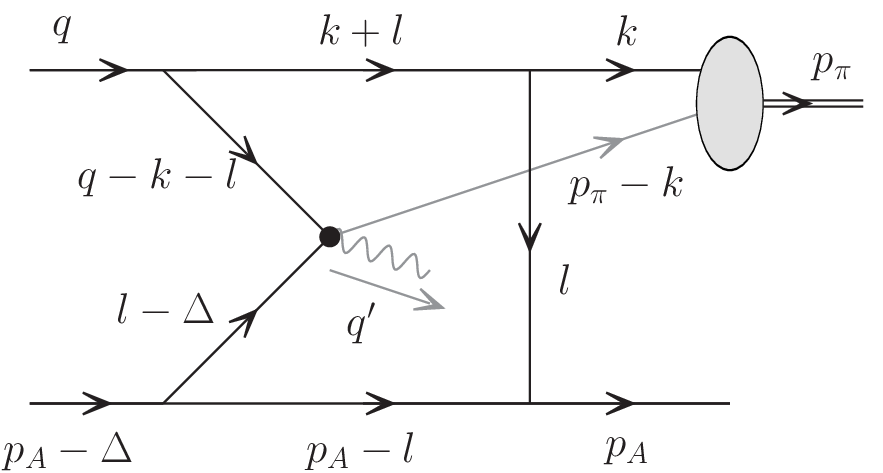}
    \caption{Simplified diagram that emphasizes the propagators that contribute to the observed collinear-to-soft Glauber pinch. The grey lines are irrelevant for the Glauber pinch analysis.}
    \label{fig:simplified-momentum-flow}
\end{figure}

As discussed earlier, the soft pinch is always present, and should be considered. In this subsection, we focus on the soft scaling for the loop momenta, given by $k \sim (\lambda, \lambda, \lambda)$ and $l \sim (\lambda, \lambda, \lambda)$. However, due to the poles of the other propagators in the amplitude, it might happen that the $+$ and/or $-$ components are pinched to be much smaller, e.g. $\lambda^2$, while the transverse component still scales as $ \lambda$. This corresponds to precisely the Glauber pinch situation.

Therefore, we start by fixing $k \sim (\lambda,\lambda,\lambda)$, and $l_{\perp}\sim \lambda$, and focus on the pole structure of $l^{\pm}$. The pinch analysis then reduces to the one-loop graph in \FIG\ref{fig:simplified-momentum-flow}. Recall that $\Delta^+ < 0$ and $q^- > 0$ kinematically. 
Then, the four relevant propagators can be written as
\begin{align}
\label{eq:pA-l}
(p_A - l)^2 + i \epsilon &= 2p_A^+ (-l^- + O(\lambda^2) + \text{sgn}(p_A^+) i\epsilon), 
\\
\label{eq:Delta-l}
(\Delta - l)^2 + i \epsilon &= - 2 \Delta^+ (l^- + O(\lambda^2) + i \epsilon),
\\
\label{eq:q-k-l}
\hspace{-0.6cm}(q-k-l)^2 + i \epsilon &= 2q^- (- l^+ + O(\lambda) +  i\epsilon),
\\
\label{eq:k-plus-l}
(k+l)^2 + i \epsilon &= 2k^- (l^+ + O(\lambda) + \text{sgn}(k^-) i\epsilon).
\end{align}
These estimates remain true also in the center-of-mass frame wrt to $q$ and $\Delta$, where $q$ is proportional to $n$, and $p_A$ and $\Delta$ are proportional to $\bar n$.

From \EQs\eqref{eq:pA-l} and \eqref{eq:Delta-l}, we thus find that $l^{-}$ is pinched to be of $O(\lambda^2)$ for $p_A^+>0$. This is also true in the collinear region. This fact alone does not necessarily imply that $l$ in pinched in the Glauber region. It is important to also confirm that $l^+$ cannot be deformed to be much larger than $\lambda$. From \EQs\eqref{eq:q-k-l} and \eqref{eq:k-plus-l}, we indeed find that $l^{+}$ is pinched\footnote{In the case where the incoming photon is virtual, the $(q-k-l)^2$ propagator is hard by definition, so $l^+$ is not pinched anymore. Hence, the corresponding Glauber pinch identified here occurs in photoproduction only.} to be of $O(\lambda)$ for $k^{-}>0$.

We have shown that $l$ is pinched to be in the Glauber region for a specific routing of the loop momentum $l$. As discussed in \SEC III E of \cite{Collins:1997sr}, to show that $l$ is truly pinched, one needs to show that for all possible routings of $l$, the pinching of $l$ persists. In the fixed order analysis in \FIG\ref{fig:explicit-2-loop}, being a two-loop graph, only two possible routings for $l$ exist (related to each other through translations of the $k$ loop momentum). In particular, the pinch in $l^-$ is independent of the routing, since it originates from propagators (\EQs\eqref{eq:pA-l} and \eqref{eq:Delta-l}) which are independent of the other loop momentum $k$. It thus suffices to show that $l^+$ is pinched in both routings. In \cite{Nabeebaccus:2023rzr}, we show explicitly that for the other routing not discussed explicitly here, $l^+$ is still pinched.

We remark that the topology of the graph in \FIG\ref{fig:simplified-momentum-flow} is actually very similar to the classic Glauber pinch that occurs in the Drell-Yan process.\footnote{We remark the well-known fact that in the inclusive Drell-Yan process, the Glauber contribution, while pinched, is canceled at the cross-section level due to unitarity.} In \FIG\ref{fig:simplified-momentum-flow}, the Glauber exchange occurs between the spectator collinear quark of the nucleon, and another \textit{soft} spectator quark joining the incoming photon with the  outgoing meson. This therefore corresponds to a $\bar{n}$-coll.~to soft Glauber, \EQ\eqref{eq:nbar-coll-to-soft-Glauber}, which is exactly the pinch that we demonstrate in this subsection. In the Drell-Yan case on the other hand, the Glauber exchange occurs between two collinear spectator partons of the two incoming nucleons, leading to the more `standard' Glauber scaling in \EQ\eqref{eq:coll-to-coll-Glauber}.

\subsection{Power counting in the Glauber region}

	We consider the scalings
 \begin{align}
     k \sim  \left(  \lambda,  \lambda , \lambda \right), \qquad l \sim  \left(  \lambda , \lambda ^2, \lambda  \right).
     \label{eq:Glauber-scaling}
 \end{align}
We simplify our analysis again by going to the light-cone gauge. This reveals the possible cancellations in generic gauges of superficially leading polarizations in the sum over graphs through WIs. Of course, the analysis can be also made in the Feynman gauge with the same conclusion \cite{Nabeebaccus:2023rzr}. Considering the replacement in \EQ\eqref{eq:lc gauge}, we select the term $- \frac{n_{\nu'} l_{\perp \nu}}{l^+} \sim \lambda^0$ of the polarization matrix corresponding to the pair of indices $\nu = \perp$ and $\nu' = +$ in \FIG\ref{fig:explicit-2-loop}. It is straightforward to check that the other terms will not give a leading contribution. Moreover, with the same argument above as in the collinear region, we may pick the $\mu$ and $\mu'$ indices to be transverse for the sake of power counting. For these terms, we have
		\begin{align}
		F_{A}^{\mu'_{\perp} +} &\sim  \lambda^4 \frac{ \lambda ^1}{ \lambda ^6}= \lambda ^{-1}\,,\quad
		F_{B} \sim  \lambda ^3 \frac{ \lambda ^3}{ \lambda ^4}= \lambda ^2\,,\nonumber\\[5pt]
		F_{H}^{ \mu_{\perp} \nu_{\perp} } &\sim  \lambda ^{2} \frac{ \lambda }{ \lambda^{3} }= \lambda ^0\;.
	\end{align}
	Hence, we find that the overall power of this contribution is $\lambda$, which is the leading power, as was claimed. 

	Finally, we note that our power counting analysis is consistent with the soft-end suppression from the pion distribution amplitude, since $ \phi_{\pi} \sim \frac{1}{f_{\pi}} F_{B} \sim  k^{-} \sim  \lambda$ (where we used $f_{\pi} \sim \lambda Q$). 
	This is discussed in more detail in \SEC VII of \cite{Nabeebaccus:2023rzr}.

\subsection{Divergence assuming collinear factorisation}

We now argue that the above Glauber pinch results in an ``endpoint-like'' divergence that appears when calculating the amplitude assuming standard collinear factorization. Strictly speaking, this divergence occurs at the combined endpoint of the pion DA $\phi_{\pi}(z)$ where $z \rightarrow 0$ and the breakpoint of the GPD $H_g(x,\xi)$ where $x \rightarrow \xi$. Notice that this corresponds to the reduced graph in \FIG\ref{fig:subset-reduced-diagrams} (b). 
Now, we have shown for the example in \FIG\ref{fig:explicit-2-loop} that there is a leading contribution from this region when $k \sim (\lambda, \lambda, \lambda)$ and $l \sim (\lambda, \lambda^2, \lambda)$. In particular, in \cite{Nabeebaccus:2023rzr} it is shown that the region where $l,k \sim (\lambda, \lambda, \lambda)$ is power-suppressed $O(\lambda^2)$. We therefore conclude that the divergence is not due to the soft region, as is usually the case with endpoint divergences.

For illustrative purposes, we note that if we were to take \FIG\ref{fig:explicit-2-loop}, and \textit{assume} collinear factorisation in terms $\phi_{\pi}(z) \sim z(1-z)$ and $H_g(x,\xi)$, we would arrive, in the limit $x \to \xi$ and $z\to 0$, at the divergent integral
\begin{align}
\int_{-1}^{1} \! \! dx \!\int _{0}^{1}\!\! dz\frac{(x-\xi) z \, H_g(x,\xi)}{ \left[  \left( x-\xi \right)+ A z - i\epsilon   \right] \!\left[(x-\xi) z + i\epsilon\right]\!\left[  x-\xi +i\epsilon   \right]   }\;,
   \label{eq:explicit-result}
\end{align}
where $A$ is a positive constant. We refer the reader to \APP A in \cite{Nabeebaccus:2023rzr} for details on the calculation.

\section{Conclusion}

We have shown that the exclusive photoproduction of a pair of $\pi^0\gamma$ with large invariant mass cannot be analyzed in a purely collinear factorization framework at the leading twist, due to a Glauber pinch for the gluon exchange channel. The pinched Glauber gluon in our analysis corresponds to one of the two active gluons that would usually participate in the hard partonic scattering level -- We have argued that it is for this reason that the amplitude diverges if one attempts to compute this contribution already at leading order and leading twist, assuming collinear factorization na\"ively. This is substantiated by the fact that the divergence only appears for those diagrams where the Glauber pinch occurs. Our results further imply that the corresponding crossed process of $\pi^0$-nucleon scattering to two photons also suffers from the same Glauber pinch. 

To save the phenomenology for the photoproduction of a $\pi^0\gamma$ pair, it is necessary to go \textit{beyond} collinear factorization - The natural approach is to introduce $k_T$-dependent distributions. We intend to address this issue in the future. However, we stress that for cases where the gluon exchange channel is forbidden, either due to electric charge conservation in the case of charged meson production, or due to C-parity conservation in the case of neutral vector meson production, the Glauber pinch discussed in this letter does not occur.

	\section{Acknowledgements}
	
	We would like to thank Renaud Boussarie, Volodya Braun, John Collins, Markus Diehl, Goran Duplan\v{c}i\'{c}, Alfred Mueller, Melih Ozcelik, Kornelija Passek-Kumeri\v{c}ki, Bernard Pire, George Sterman, Iain Stewart, Jianwei Qiu, Zhite Yu  for many useful discussions. This work was supported by the GLUODYNAMICS project funded by the ``P2IO LabEx (ANR-10-LABEX-0038)'' in the framework ``Investissements d’Avenir'' (ANR-11-IDEX-0003-01) managed by the Agence Nationale de la Recherche (ANR), France. This work was also supported in part by the European Union’s Horizon 2020 research and innovation program under Grant Agreements No. 824093 (Strong2020). This work was partly supported by the Science and Technologies Facilities Council (STFC) under grant  ST/X00077X/1. This project has also received funding from the French Agence Nationale de la Recherche (ANR) via the grant ANR-20-CE31-0015 (``PrecisOnium'')  and was also partly supported by the French CNRS via the COPIN-IN2P3 bilateral agreement. J.S. was supported by the Research Unit FOR2926 under grant 409651613, by the U.S. Department of Energy through Contract No. DE-SC0012704 and Laboratory Directed Research and Development (LDRD) funds from Brookhaven Science Associates.

	\bibliographystyle{utphys}
	
	\bibliography{masterrefs.bib}

\end{document}